\documentclass[]{interact}

\usepackage{epstopdf}
\usepackage[caption=false]{subfig}
\usepackage{amsmath, amsfonts, amssymb} 
\usepackage[english]{babel}
\usepackage{color} 
\usepackage{graphicx} 
\usepackage{url} 
\usepackage{bm} 
\usepackage{multirow}
\usepackage{booktabs}
\usepackage{epstopdf}
\usepackage{epsfig}
\usepackage{algorithm}
\usepackage{algorithmicx}
\usepackage{titlesec}
\usepackage{booktabs}
\usepackage{algpseudocode}
\usepackage{amsmath}
\usepackage{xcolor}
\usepackage{tikz}
\usepackage{amssymb}
\renewcommand{\algorithmicrequire}{ \textbf{Input:}}
\renewcommand{\algorithmicensure}{ \textbf{Initialize:}}

\usepackage{multicol}
\usepackage{enumerate}
\usepackage{hyperref}
\usepackage{algorithm,algpseudocode,float}
\usepackage{lipsum}
\usepackage{amsmath}
\usepackage[utf8x]{inputenc}
\usepackage[titletoc]{appendix}
\usepackage{mathrsfs}
\usepackage{amsmath,amsthm,amsfonts,amssymb,amscd}

\makeatletter
\newenvironment{breakablealgorithm}
{
\begin{center}
\refstepcounter{algorithm}
\hrule height.8pt depth0pt \kern2pt
\renewcommand{\caption}[2][\relax]{
{\raggedright\textbf{\ALG@name~\thealgorithm} ##2\par}%
\ifx\relax##1\relax 
\addcontentsline{loa}{algorithm}{\protect\numberline{\thealgorithm}##2}%
\else 
\addcontentsline{loa}{algorithm}{\protect\numberline{\thealgorithm}##1}%
\fi
\kern2pt\hrule\kern2pt
}
}{
\kern2pt\hrule\relax
\end{center}
}
\makeatother

\theoremstyle{plain}

\theoremstyle{definition}

\theoremstyle{remark}

\begin{document}


\title{Design of High-Frequency Trading Algorithm Based on Machine Learning}

\author{
\name{Boyue Fang\textsuperscript{a}\thanks{Email: byfang16@fudan.edu.cn} and Yutong Feng\textsuperscript{b}}
\affil{\textsuperscript{a}School of Mathematical Sciences, Fudan University ; \textsuperscript{b}Electronics Science and Technology, Fudan University}
}

\maketitle
\date={}
\begin{abstract}
Based on iterative optimization and activation function in deep learning, we proposed a new analytical framework of high-frequency trading information, that reduced structural loss in the assembly of Volume-synchronized probability of Informed Trading ($VPIN$), Generalized Autoregressive Conditional Heteroscedasticity (GARCH) and Support Vector Machine (SVM) to make full use of the order book information. Amongst the return acquisition procedure in market-making transactions, uncovering the relationship between discrete dimensional data from the projection of high-dimensional time-series would significantly improve the model effect. $VPIN$ would prejudge market liquidity, and this effectiveness backtested with CSI300 futures return.
\end{abstract}


\section{Introduction}

How to mine the information of the order book has always been a concern of market traders, especially the high-frequency market traders. Many conventional models researched over one dimension of information like price change and volatility, but these models only employed part of the order book information. This article composes an optimizer acting on GARCH and $VPIN$ throughout the activation SVM layer. In the CSI300 futures market, the performance of this method is better than using one-dimensional data.

Market makers use trading techniques to benefit from promoting the trading of specific securities. High-frequency trading (HFT) structure has a significant impact on the profit mechanism of high-frequency traders(HFTs). Álvaro Cartea \cite{2} divided traders into ordinary traders, informed traders and market makers, and statistical arbitragers that, in the usual sense, were considered to be the informed trader. Market makers are more conservative than other traders. Because their strategies used adaptive trading techniques in the market, they can only gain small benefits in a single transaction, so it may take several thriving market making actions to make up for one loss caused by market volatility.

Liquidity traders mainly purpose on medium and long-term investments, so they will participate in transactions by formulating strategies based on the business activities of securities that they hold. Most of the profit of such trading strategies comes from portfolio management and risk-return, and there is little short-term price information that returns by taking risks, which is exceeding the information provided by the size of their positions. Thus, we can treat transactions of this type of traders as noise transactions from the perspective of HFT. Because informed traders have some undisclosed information, they get excess returns privileged by information advantage, but these returns come from losses from market makers and liquidity traders. So the number of informed traders will affect the profitability of market-makers.

Therefore, market makers want to measure the effect of reducing their losses. Grossman \cite{15}and Easley \cite{10} discussed that liquidity traders provide liquidity premiums to compensate for the risk during market makers' holding period. By maximizing the utility function, they provided the principle that the market makers solve the liquidity demand of the liquid trader and place the risk premium. Other models also measured informed trading probability, such as the non-synthetic risk metrics used by Benston \cite{4} and the delegated imbalance indicators used by Aktas \cite{1}, but these methods are indirectly measuring the probability of informed trading. The EKOP model proposed by Easley \cite{11} provided a new idea for directly measuring the degree of informed trading. The method calculated the expected number of orders received to get the informed trading probability (PIN), and Easley \cite{12} furthered $VPIN$ by taking the informed transaction rate into account. Cheung \cite{6} applied the $VPIN$ method to the mandatory recall event(MCE) of Hong Kong stocks. The $VPIN$ value observation increased significantly before the incident occurred; in other words, the effect approached the liquidity risk warning. Liu \cite{22} used the $VPIN$ method in China Commodity futures markets, and the $VPIN$ method also fitted for the commodity futures market. Besides, Lu \cite{18} tested various forms of heterogeneity in the framework of nonparametric varying-coefficient models; the idea is similar to the $VPIN$ method. High-frequency trading involves massive data, which consists of a growing number of heterogeneous subpopulations. The $VPIN$ method uses homogeneity testing to find the test statistic aggregated over all sub-populations. In a word, after obtaining liquidity risk warning in $VPIN$, market makers can get better earnings performance in specific trading markets. Low \cite{17} found support for the application of the BV-$VPIN$ model in international equity markets as a risk monitoring and management tool for portfolio managers and regulators.Yan(2018)\cite{24} compared the DPIN method and $VPIN$ method in predicting the future price in the CSI300 index futures market. They found that DPIN could effectively capture price information in the future index markets in China.

However, the $VPIN$ method only used part of the information of the securities, the information of the higher dimensional time series was not fully excavated. Therefore,  we introduced the GARCH model for analytics on the logarithmic rate of return.

Engle \cite{14} proposed the autoregressive conditional heteroscedasticity (ARCH) model. He applied it to study the volatility of financial time series. Bellerslev \cite{5} proposed the GARCH model based on the ARCH model to fit heteroscedastic functions with long-term memory. Wang \cite{21} applied the GARCH model to the Xinhua FTSE A50 stock index futures. He found that the stock index futures made the volatility of China's stock market larger. Deng \cite{9} used the GARCH-VaR model to forecast the risk of investors portfolios. The experimental results on the transaction data of the futures portfolio showed the effectiveness of their proposed approach. Choudhry \cite{7} used the GARCH model to forecast the daily dynamic hedge ratios in emerging European stock futures markets. By comparing the different models, they found the GARCH-BEKK and GARCH-GJR had excellent performance.

Reconciling the models above over different dimensions needs the understanding of the relationships between logical iterative optimization methods. Therefore, this article uses the Granger causality test to test the original hypothesis that the value of $VPIN$ is not the Granger cause of the logarithmic rate of return of CSI300 futures. The result showed that after a two-stop lag, the test result of the P-value about the value of $VPIN$ is not the value of the Granger cause of the logarithmic rate of return of CSI(Chinese stock index)300 futures is 0.02. If the significance level is 5\%, rejected the original hypothesis. Therefore, $VPIN$ is the Granger causality of the logarithmic rate of return of CSI300 futures.

Moreover, when $VPIN$ is large enough, informed traders are flooding the market, market-makers need to turn into a robust strategy to prevent massive losses. When $VPIN$ is reasonable, they can change into aggressive market-making strategies. Lopez \cite{16} used machine learning to analyze the critical and reductive importance of some of the best-known market microstructural features. Machine learning skills helped them to analyze tick-data history in 10 years. In order to balance the combination of these two models, this article uses the support vector machine (SVM) to increase the forecasting ability of strategies. The effectiveness of this optimization methodology was backtested with observations of the CSI 300 futures and illustrated a more profound excess return.

\section{Background}
This section provides a necessary background for market making. Liquidity traders provide a liquidity premium to compensate for the price risk of market makers holding securities. Here, we use a simple model to illustrate the problem of financial market liquidity premium from the perspective of microstructure. Assume market makers have certain initial assets to purchase securities, $t\in\{1,2,3\}$. Liquidity trader 1 ($LT_1$) who holds I unit securities is willing to sell securities at $t=1$, and liquidity trader 2 ($LT_2$) is willing to buy securities at $t=2$, where transaction cost was assumed zero to emphasize the securities price change.

The value of the securities at $t=3$ is $S_3=\mu+\xi_2+\xi_3$, where $\mu$ is a constant, $\xi_2,\xi_3 \thicksim IIDN(0,\sigma^2)$. $\xi_2$ reflects the change from $t_1$ to $t_2$, $\xi_3$ reflects the change from $t_2$ to $t_3$. All traders are risk averse, the utility function $U(X)=-e^{-\gamma X}$, where $\gamma>0$ is the risk aversion parameter, which reflects the loss of utility for risk. So they are aim to maximize the $E[U(X_3)]$.

Market makers ($MM$), $LT_1$, $LT_2$ holds the securities in quantities of $q_1^{MM}$, $q_1^{LT_1}$, $q_1^{LT_2}$ respectively at $t=2$. Shorthand for $q_1^j$, where $j\in\{MM,LT_1,LT_2\}$ (eg: $q_t^j=2$ means the trader $j$ holds 2 unit securities at $t+1$), traders are aim to maximize expected returns, that is:

\begin{equation*}
\max_{q_2^j}\mathbb{E}[U(X_3^j)|\xi_2],
\end{equation*}
where $X_3^j=X_2^j+q_2^j S_3$, $X_2^j+q_2^j S_2=X_1^j+q_1^j S_2$.

And $\xi_2,\xi_3 \thicksim IIDN(0,\sigma^2)$, so
\begin{equation*}
\mathbb{E}[U(X_3^j)|\xi_2]=-exp{\{-\gamma(X_2^j+q_2^j\mathbb{E}[S_3|\xi_2])+\frac{1}{2} \gamma^2 {(q_2^j)}^2 \sigma^2\}}.
\end{equation*}

So for all traders, the optimal number of securities held is:

\begin{equation*}
q_2^{j,*}=\frac{\mathbb{E}[S_3|\xi_2]-S_2}{\gamma\sigma^2}.
\end{equation*}

At $t=2$, supply is equal to demand, so

\begin{equation*}
nq_1^{MM}+q_1^{LT_1}+q_1^{LT_2}=nq_2^{MM}+q_2^{LT_1}+q_2^{LT_2}.
\end{equation*}

As we evidence above, $q_2^j$ are equal for all $j$, so
\begin{equation*}
nq_1^{MM}+q_1^{LT_1}+q_1^{LT_2}=(n+2)\frac {\mathbb{E}[S_3|\xi_2]-S_2}{\gamma\sigma^2}.
\end{equation*}

The total value of all securities at $t=1$ is the value of the securities sold by $LT_1$, so $nq_1^{MM}+q_1^{LT_1}+q_1^{LT_2}=i+q_1^{LT_2}=i-i=0$. At $t=2$, $S_2=\mathbb{E}[S_3]=\mu+\xi_2+\mathbb{E}[\xi_3]=\mu+\xi_2$, so $q_2^j=0$. The economic meaning of this formula is that when $t=3$, no one wants to hold risky securities.

When $t=1$, there are only $n$ market makers and $LT_1$ in the market, which means that no matter what they do, the future market is valid, and they would not hold securities at $t=3$, so $X_3=X_2$. and the portfolio strategy goal is

\begin{equation*}
\max_{q}\mathbb{E}[U(X_2^j)],
\end{equation*}
where $X_2^j=X_1^j+q_1^j S_1$, $X_1^j+q_1^j S_1=X_0^j+q_0^j S_1$.

Repeating the analysis like $t=1$, For all traders at this time, the optimal decision is $q_1^{j,*}=\frac{E[S_2]-S_1}{\gamma\sigma^2 }$. Similarly, when $t=1$, the supply and demand of the securities are equal, so
\begin{equation*}
n q_0^{MM}+q_0^{LT_1}=n q_1^{MM}+q_1^{LT_1},
\end{equation*}
where $ q_0^{LT_1}=i$ which is the share that $LT_1$ wants to sell), $q_0^{MM}=0$, so $S_1=\mu-\gamma\sigma^2 \frac{ i}{n+1}$. The following figure shows this process.

\begin{figure}[H]
\centering
\includegraphics[width=1\textwidth]{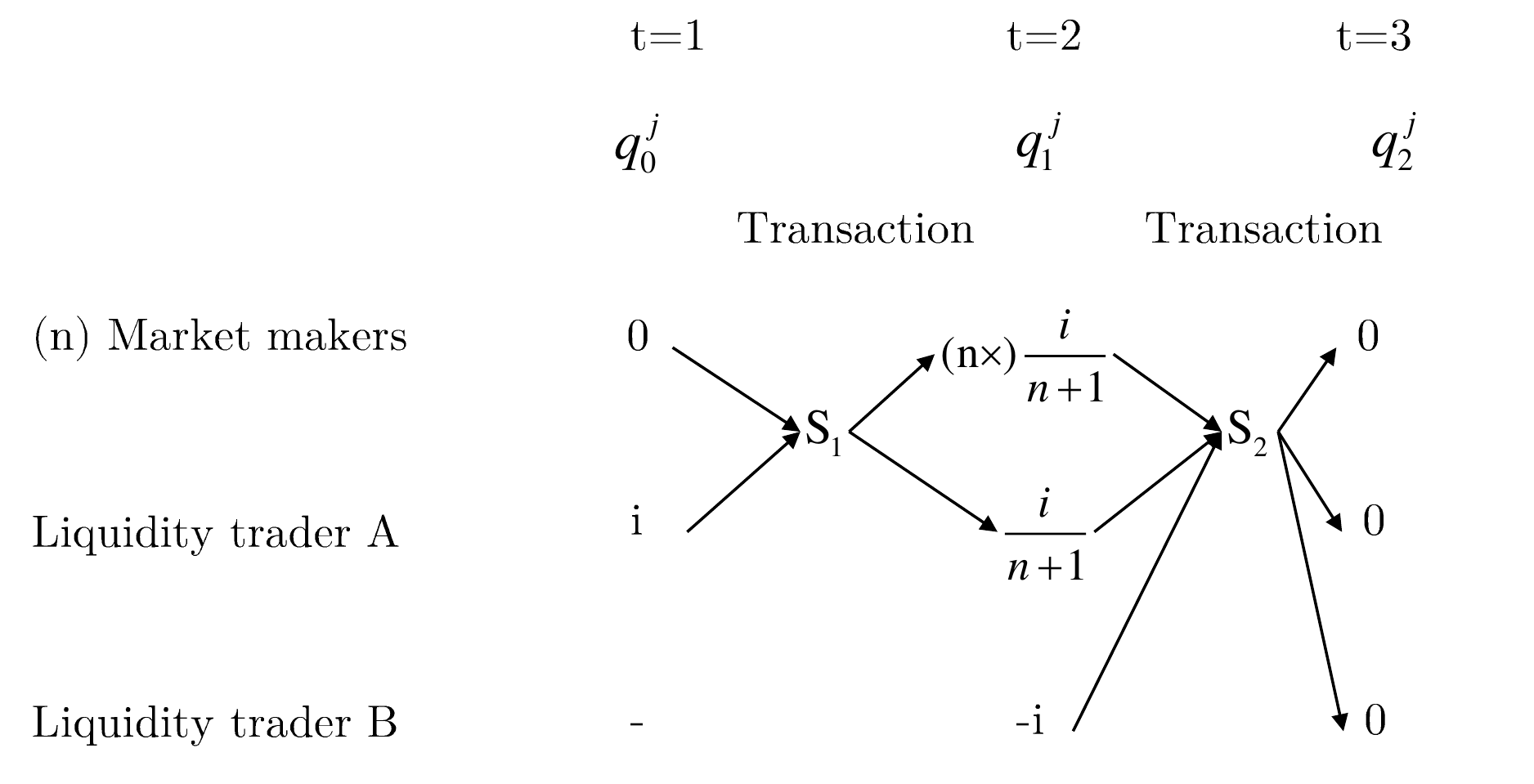}
\caption{Transaction process}
\end{figure}

From this analysis, this paper explains how the market solves the liquidity requirement of $LT_1$. When the market makers get sufficient liquidity compensation, they will accept the securities, and $LT_1$ is price sensitive, so if he accepts liquidity compensation, he will not sell all securities at once. In equilibrium, $LT_1$ and market makers hold $q_1^{j,*}$ unit securities; ultimately, the transaction price is lower than the valid price. Furthermore, the spread $|S_1-\mu|=\gamma\sigma^2 \frac{i}{n+1}$ shows that as the competition between market makers increases (n increases), the risk premium decreases, which is related to the Chinese futures markets. It also shows the importance of the information differences. This article will use the $VPIN$ method later to measure the impact of informed trading.

High-frequency trading will bring lots of trading volume to the market, but each transaction will affect the current equilibrium price. So we need to focus on how to minimize the expectation of the total execution cost. In order to solve this stochastic control problem, This article uses the dynamic programming theorem and the nonlinear partial differential Hamilton-Jacobi-Bellman equation.

\section{Theoretical Concept}
\subsection{Generalized Autoregressive Conditional Heteroscedasticity(GARCH)}
the structure of the GARCH model is as follows:

\begin{equation*}
\left\{
\begin{aligned}
&\varepsilon _t=e_t \sqrt{h_t} \\
&y_t=\gamma x_t'+\varepsilon _t \\
&h_t=\sum _{j=1}^q \gamma _j h_{t-j}+\sum _{i=1}^p \alpha _i \varepsilon _{t-i}^2+\alpha _0,
\end{aligned}
\right.
\end{equation*}
where $e_t \thicksim IIDN(0,1)$, $\alpha_0$ is constant and $\alpha_0>0$, $\alpha_i\geqslant0$,$\gamma_i\geqslant0$,$\sum_{i=1}^{max(p,q)}(\alpha_i+\gamma_i)<1$, $\varepsilon$ is the residual term, $\varepsilon_t$ obey the GARCH($p$,$q$) process. $h_{t-j}$ is the conditional variance term, which describe the memory effect of volatility. And it is necessary to examine the ARCH effect before using the GARCH model.

\subsection{Volume-synchronized Probability of Informed Trading($VPIN$)}
$VPIN$ method uses the bulk volume classification to divide the transaction volume in the sample period into 50 parts, define each as a basket, and denote the volume of the transaction in the basket as $VBS$. The basket is filled in: from the beginning of the transaction. If the transaction volume exceeds the current basket, then the remaining portion will be calculated in the next basket. Finally, there will be a list of baskets. For any basket $\tau$, buy and sell transaction volume is:

\begin{equation*}
V_\tau^B=\sum_{i=t(\tau-1)+1}^{t(\tau)} {V_i \Phi(\frac{\Delta P_i}{\sigma_{\Delta P}}) }
\end{equation*}
\begin{equation*}
V_\tau^S=\sum_{i=t(\tau-1)+1}^{t(\tau)} {V_i [1-\Phi(\frac{\Delta P_i}{\sigma_{\Delta P}})] }=VBS-V_\tau^B,
\end{equation*}
where $V_i $ is the volume of time i, $i\in [t(\tau-1)+1,t(\tau)]$, $t(\tau)$ represents the time of the last transaction in the $\tau$ trading basket, $\Phi$ is the distribution function of the standard normal distribution, $\Delta P_i$ is the price change for the time interval, and $\sigma_{\Delta P}$ is the standard deviation of price change for all samples. In economics, this formula means: For a certain time interval, if $V_\tau^B=V_\tau^S$, the motivation for buying and selling is consistent, and the influence of the information is just balanced; if $V_\tau^B>V_\tau^S$, then the buying motivation is greater; if $ V_\tau^B <V_\tau^S$, then the selling motivation is greater. Note that this rough estimation would not interfere with the measure of market liquidity affection, as the ultimate goal is not to seek accurate buying and selling volume, and such analytics would detect the inconsistency between the buy and sell orders.

The $VPIN$ also assumed that the price affected by the information event is $\alpha$. Therefore, the probability of bad news is $\delta$, the bid price will be $-\mu+\varepsilon$ and the offer price will be $\varepsilon$; and the probability of good news is $1-\delta$, the bid price will be $ \varepsilon$, and the offer price will be $\mu+\varepsilon$. If there is no such information, the bid and ask price will be $\varepsilon$. The situation illustrated in table 1. (The letters in parentheses indicate the probability of occurrence)

\begin{table}[H]
\centering
\begin{tabular}{@{}cccc@{}}
\toprule
& \begin{tabular}[c]{@{}l@{}}Type of\\ information\end{tabular} & \begin{tabular}[c]{@{}l@{}}bid\\ price\end{tabular} & \begin{tabular}[c]{@{}l@{}}ask\\ price\end{tabular} \\ \midrule
\multirow{2}{*}{\begin{tabular}[c]{@{}l@{}}information\\ ($\alpha$)\end{tabular}} & \begin{tabular}[c]{@{}l@{}}good news\\ (1-$\delta$)\end{tabular} & $\varepsilon$ & $\mu+\varepsilon $ \\
& \begin{tabular}[c]{@{}l@{}}bad news\\ ($\delta$)\end{tabular} & $-\mu+\varepsilon $ & $\varepsilon $ \\
\begin{tabular}[c]{@{}l@{}}no information\\ ($1-\alpha$)\end{tabular} & & $\varepsilon$ &$ \varepsilon $ \\ \bottomrule
\end{tabular}
\caption{The effect of information on prices}
\end{table}

The imbalance of each trading basket is $OI=|V_\tau^B-V_\tau^S|$, expectation is $\mathbb{E}[OI]\approx\alpha \mu$, total transaction volume arrival rate is $\mathbb{E}[VBS]=\alpha \mu+2\varepsilon$, the trading volume of expectations brought by good news is $\alpha (1-\delta)(2\varepsilon+\mu)$, the trading volume of expectations brought by bad news is $\alpha \delta(2\varepsilon+\mu)$, the trading volume of expectations is $2(1-\alpha)\varepsilon$ if there is no such information.

The value of $VPIN$ can obtain by following formula:
\begin{equation*}
VPIN=\frac{\alpha \mu}{\alpha \mu+2\varepsilon}=\frac{\alpha \mu}{V}\approx\frac{\sum_{\tau=1}^n |V_\tau^B-V_\tau^S|}{nV}.
\end{equation*}

Calculations of $VPIN$ would give an estimate of the proportion of informed traders in the market.

However, $VPIN$ does not work well in high-frequency trading, because the $VPIN$ method combines the ask volume and the bid volume in the market, then use the difference between them to predict the volatility changing later. For example, more pending orders on the bid side than the offer side would forward an upstream. Nevertheless, in high-frequency trading, it is hard to balance the number of baskets and forecasting ability. If the number of $VPIN$ baskets is too hefty, it may lead to a significant decline in the overall forecasting ability. If the number of baskets is too small, it may miss many trading opportunities. So after combining the HAR-RV model, it is a good idea to use different scales volatility to hedge.
\subsection{Support Vector Machine (SVM)}
The support vector machine is one of the most robust and accurate methods in classical data mining algorithms, and this is one of the reasons why we choose it. In high-frequency trading, both the rate of return and the efficiency of the algorithm are crucial.

Consider the linear separability problem, SVM hopes to find a hyperplane with the largest interval to separate the two types of samples, so that $\textbf w$ and $b$ respectively represent the weight vector and the optimal hyperplane offset, then the hyperplane can be defined through $\textbf w^T\textbf x+b=0$, the distance from the sample to the optimal hyperplane is $r=\frac{g(\textbf x)}{||\textbf w||}$, where $g(\textbf x)=\textbf w^T\textbf x+b$ is the functional interval of $\textbf x$ that $\textbf w$ and $b$ had been given. Through the training set, $\textbf w$ and $b$ can be given by minimum $r$. That is, for a given training set $ \left\{\textbf x_i,y_i\right\}_{i \in [ 1,n]} \in \mathbb{R}^m \times \left\{ {\pm1}\right\}$, it has:

\begin{equation*}
\left\{
\begin{aligned}
\textbf w^T \textbf x_i \ge1 \;\;y_i=+1\\
\textbf w^T \textbf x_i \le1 \;\;y_i=-1.
\end{aligned}
\right.
\end{equation*}

Then by using the Sequential Minimal Optimization(SMO) algorithm, the optimal solution of $\textbf w$ and $b$ can be given.

For the indivisible linear problem, employing soft interval optimization or kernel technique would map the original space into high-dimensional space and then transforms the problem into a linear separable problem in high-dimensional space. The radial basis kernel function is a suitable choice for this problem. The function form is $K(\textbf x_1,\textbf x_2)=e^{-\frac{||\textbf x_1-\textbf x_2||^2}{2\sigma^ 2}}$, which represent a set of functions that map the original space to the infinite-dimensional space. This kernel appeared to have better generalization capability and is more capable of high-frequency trading with huge data volume. In the next section that revised the GARCH model with SVM, the performance improved significantly.

\section{Data Inspection Based on SCI300 Futures}
\subsection{$VPIN$}
Through the $VPIN$ theoretical concept, this paper went through the $VPIN$ value from October 8, 2015, to September 30, 2018 \footnote{IF.CFE data\\data sources: Wind}, and combine the following figures with the CSI 300 futures changes:

\begin{figure}[H]
\centering
\includegraphics[width=1\textwidth]{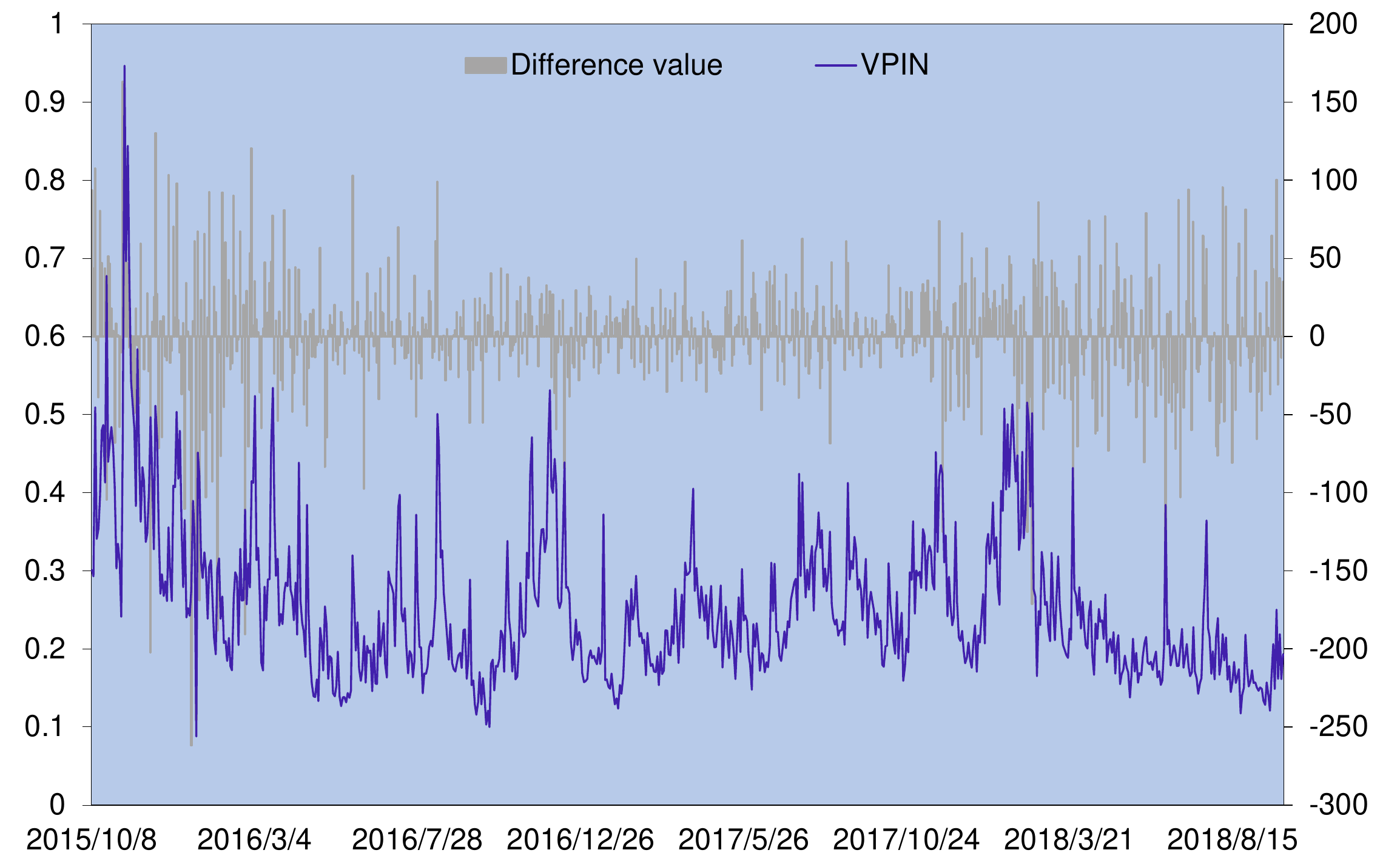}
\caption{$VPIN$}
\end{figure}

Trending the value of the futures price, it oscillates more significant when $VPIN$ becomes larger. After the stability testing, it is not difficult to know that the two time-series are stable. The Granger test can show that $VPIN$ is useful to predict the fluctuation of futures prices. For the two-step lag, under the 0.05 significance level, the $VPIN$ value is the Granger causality for the logarithmic yield of stock index futures.

\begin{table}[H]
\centering
\begin{tabular}{@{}lllll@{}}
\toprule
Null Hypothesis: & & Obs & F-Statistic & Prob. \\ \midrule
\begin{tabular}[c]{@{}l@{}}Y does not Granger\\ Cause X\end{tabular} & & 730 & 0.19412 & 0.8236 \\
\begin{tabular}[c]{@{}l@{}}X does not Granger\\ Cause Y\end{tabular} & & & 3.82724 & 0.0222 \\ \bottomrule
\end{tabular}
\caption{The result of Granger Causality Test}
\end{table}

So, in the case of a larger $VPIN$  value, it indicates that the informed trader is flooding the market. High-frequency market makers need to trade in a more preservative way to prevent significant losses. On the contrary, they can consider more aggressive market-making strategies.

\subsection{GARCH}
Again
This paper uses the CSI300 futures data from January 1, 2018, to October 1, 2018\footnote{IF.CFE data\\data sources: Wind}, the time series are its logarithmic rate of return.

\begin{figure}[H]
\centering
\includegraphics[width=1\textwidth]{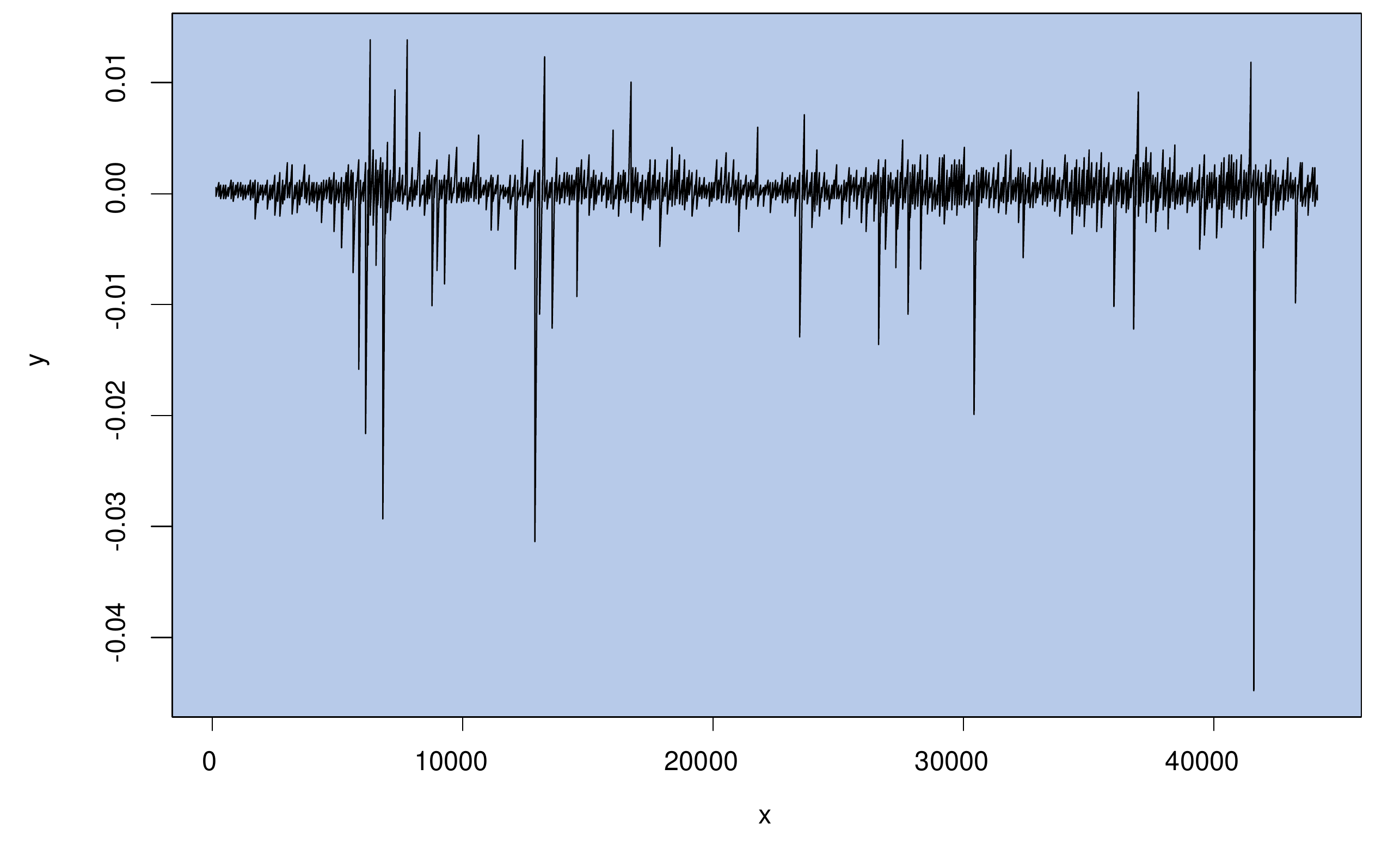}
\caption{logarithmic rate of return time series}
\end{figure}

\begin{enumerate}[1)]

\item The time series average is $-3.69\times10^{-6}$, the standard deviation is $6.32\times10^{-4}$, the skewness is -7.27, and the kurtosis is 355.63, which is much higher than the normal distribution. The sequence has the characteristics of leptokurtosis and fat-tail. The Jarque-Bera statistic is $2.30\times10^8$, and the P-value is 0.00, rejecting the assumption that the sequence obeys the normal distribution.
\item Consider the stationariness of the sequence. The station's stationarity means that the mean, variance, and autocovariance are independent of time t. The ADF test shows that the t statistic value is -101.31; the corresponding P-value is 0.0001, indicating that the logarithm rate of return time series is stable.
\item With testings on sequential and partial autocorrelation, no significant correlation in the time series was identified.
\item In order to analyze whether the time series has an ARCH effect, each number in the time series is de-equalized and then squared. The updated time series have autocorrelation (see Appendix), so there is an ARCH effect. Then consider the GARCH (1,1), GARCH (1,2) GARCH (2,1) model, respectively. All coefficients of the three models pass the residual test; for example, in the GARCH (1,1) model, $\alpha+\beta$ value is 0.93, which is close to and less than 1. In other words, the process of GARCH (1,1) is more stable, and the influence of fluctuation is gradually decaying under a slow attenuation, indicating that the market has a strong memory.
\item The TGARCH model illustrated a positive leverage effect in the market. In other words, the impact of bad news on volatility is more significant than the good news. (leverage effect can be simply understood as there is a negative correlation between current earnings and future fluctuations.) When income increases, the volatility decreases, and vice versa. The insignificance of the ARCH-M test coefficient indicated that there is no ARCH-M process.

\end{enumerate}

\section{Study Design}
The above discusses the initial warning of liquidity risk and the study of the volatility of high-frequency data. This paper will review the implementation strategy of high-frequency trading. In general, high-frequency trading is more profitable by arbitrage. However, it also leads to it is more difficult to make high-frequency trading profits, so it is necessary to examine new directions. Furthermore, we use machine learning to combine the existing technology to innovate the model to market-making.

As mentioned above, we have obtained some predictions on the liquidity risk and analyzed the volatility of the logarithmic rate of return time series. We naturally hope to utilize the obtained conclusions to guide the design of the high-frequency trading algorithms. The statistical arbitrage aims to judge the arbitrage opportunity according to the established model. Because of the lag and asymmetry of information, there are some arbitrage opportunities in the market. According to the established correlation, we can find the law of individual securities prices by using statistical tools, to establish a position profit in the direction of the mean return. This process is similar to the idea of using the GARCH model. In other words, arbitrage and market-making are not entirely separate.

First, we use the GARCH model to make the most fundamental predictions and determine the trading direction, as discussed in the GARCH section. The key here is the choice of thresholds. The threshold is determined based on the principle of volatility aggregation, and our goal is to maximize the yield. So firstly, we need to determine the time window. This paper chooses the current time (excluding the current time) to 1 hour before, the initial threshold $\delta_1\in [0.02,2]$, the step size is 0.02, and then take the value step by step, finally, choose the threshold that maximizes the yield one second before as the current threshold.

\renewcommand{\algorithmicrequire}{\textbf{Input:}}
\renewcommand{\algorithmicensure}{\textbf{Output:}}
\begin{breakablealgorithm}
\caption{GARCH Section}
\begin{algorithmic}[1] 
\Require {logarithmic rate of return time series}
\Ensure {determine the direction of trading for futures}
\State Input {logarithmic rate of return time series}
\State Use GARCH(1,1) model to get predicted value
\State Output\;{variance}
\For {$\delta_1 \in [0.02,2] $, step \;size = 0.02}
\For {time $\in {[now-1hour,now)}$}
\If {predicted value $> \delta_1$}
\State buy IF in bid1 price
\ElsIf {predicted value $< -\delta_1$}
\State ask IF in ask1 price
\State \Return{ the rate of return}
\EndIf
\State \Return {$\arg\max_{\delta_1}\left\{rate\; of\; return \right\}$}
\EndFor
\If {predicted\; value $> \delta_1$}
\State buy IF in buy1 price
\ElsIf {predicted value $< -\delta_1$}
\State ask IF in ask1 price
\EndIf
\State \Return{the direction of trading for futures}
\EndFor
\end{algorithmic}
\end{breakablealgorithm}

Then, as we discussed in the $VPIN$ section, the Granger causality test between the $VPIN$ value and the logarithmic rate of return time series shows that the $VPIN$ method is useful for improving the high-frequency trading algorithms. Moreover, the GARCH model does not use the trading volume information, so combining the $VPIN$ method can help to improve the stability and profitability of the high-frequency trading model based on the GARCH model.

We consider the specific use of $VPIN$ in combination with GARCH. First, the algorithm needs to compare the current $VPIN$ value with the threshold $\delta_2$ and $\delta_3$, which depends on a similar way as threshold $delta_1$ in GARCH model and delay the two "basket" time to adjust the trading threshold of the GARCH model. Therefore, to obtain the threshold, we first select the previous trading day, and today until the current time (excluding the current time), the threshold is in the range of [0,1]. The stochastic gradient descent algorithm gives the $VPIN$ delay. The most appropriate value (i.e., if $VPIN$ exceeds the $\delta_2$ threshold two "basket" times before, the price does have large fluctuations, if $VPIN$ is less than the $\delta_3$ threshold two "basket" times before, the price does have few fluctuations. "basket" time means the time interval between start filling the basket to fill the basket, and it depends on the market trading.) for the actual fluctuations is used as the threshold two "basket" time later. Moreover, if $VPIN$ is larger than the $\delta_2$ threshold, the GARCH threshold will be revised to the arithmetic mean of the standard threshold and the maximum threshold for the current day. If $VPIN$ is less than the $\delta_3$ threshold, the threshold of the GARCH model will be revised to the arithmetic mean of the standard threshold and the minimum threshold for the current day.

\begin{breakablealgorithm}
\caption{$VPIN$ Section}
\begin{algorithmic}[1] 
\Require logarithmic rate of return time series
\Ensure determine the direction of trading for futures
\State Import {$VPIN$ model}
\State Import {stochastic gradient descent model} as stoc
\For {$\delta_2,\delta_3 \in [0,1]$}
\For {time $\in {[now-1hour,now)}$}
\State use stoc to determine $\delta_2$,$\delta_3$ which fit best that when $VPIN>\delta_2$,the fluctuation(after\; 2\; periods)$>$0.15\%,and when $VPIN<\delta_3$,the fluctuation(after\; 2\; periods)$<$0.05\%
\EndFor
\State
\begin{equation*}
\delta_1=\left\{
\begin{aligned}
\frac{\delta_1+\max_{t \in nowaday}{\delta_1}}{2}\;\;\;VPIN>\delta_2\\
\frac{\delta_1+\min_{t \in nowaday}{\delta_1}}{2}\;\;\;VPIN<\delta_3
\end{aligned}
\right.
\end{equation*}
\EndFor
\end{algorithmic}
\end{breakablealgorithm}

SVM: Take the RBF kernel function, $\sigma $ parameter takes 0.0001, train the predicted value in the GARCH model of the first 30 trading days, and the IF.CFE price to correct the GARCH model (i.e., if the SVM predicts that the trading of IF.CFE will make some loss, but in the GARCH model(or GARCH+$VPIN$ model) it suggests to trade, then cancel this trading).

\begin{breakablealgorithm}
\caption{SVM Section}
\begin{algorithmic}[1] 
\Require logarithmic rate of return time series
\Ensure determine the direction of trading for futures
\State Use RBF kernel, $\sigma=0.0001$
\State The training set includes the predicted value by using the GARCH model and the IF.CFE value for the latest 30 trading days.
\If {SVM predicted value is -1(which means the price will fall)}
\If {in GARCH model we will trade in bid1 price}
\State Interrupt the trade
\EndIf
\EndIf
\If {SVM predicted value is 1(which means the price will rise)}
\If {in GARCH model we will trade in ask1 price}
\State Interrupt the trade
\EndIf
\EndIf
\end{algorithmic}
\end{breakablealgorithm}

Finally, review the number of securities held and the position of stop loss. As the analysis in $VPIN$, $VPIN$ is essentially a measure of the impact of informed trading, so the addition of $VPIN$ will affect the choice of position. This article used 10\% of currently available funds in each trading if the $VPIN$ method not added. After adding the $VPIN$ method, adjust the position similar to the GARCH threshold. If the $VPIN$ is greater than the threshold, then reduce the investment, and vice versa. The feature of intraday transactions would trap the stop loss within two standard deviations.

Next, analyze the performance of whether to combine the GARCH model and the $VPIN$ method and SVM.

Although the use of 500 milliseconds of transaction data is more in line with high-frequency trading, the strategy tests 500 milliseconds of data and one minute of data, respectively. It shows the change is very little, indicating that the strategy is not sensitive to ultra-high frequency(UHF) data. This paper will explain this phenomenon later and propose new improvements.

The initial capital of the strategy is set at 10 million, the margin is 25\%, and the transaction fee is 6.87\%\%. The backtest interval is from January 1, 2016, to September 30, 2018. Backtesting during the whole time in minute-level data. The figures[3-8] shows the backtesting performance.

\begin{figure*}
\centering
\includegraphics[width=1\textwidth]{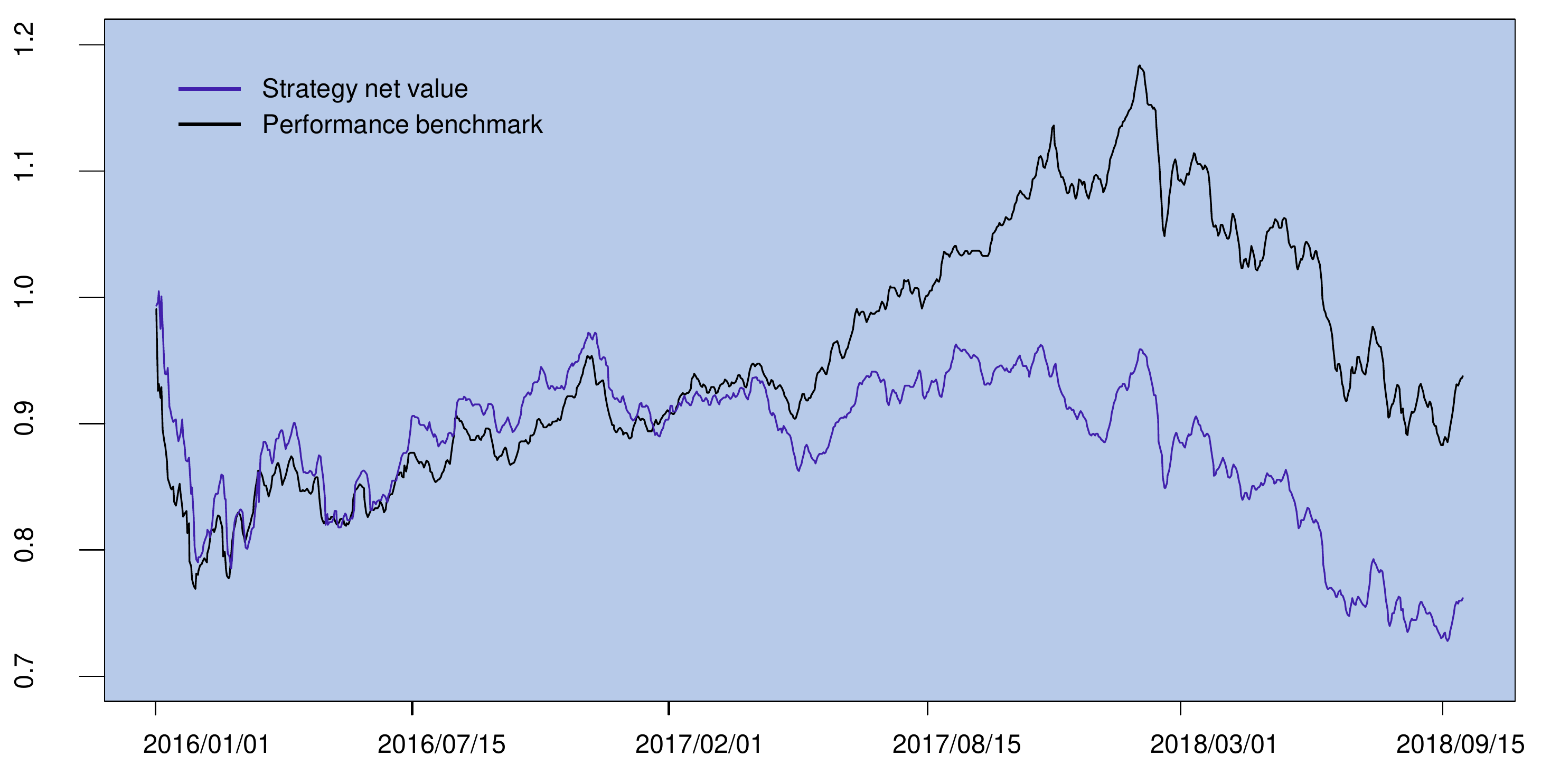}
\caption{Using GARCH model(without $VPIN$, SVM)}
\label{imgGARCH only}
\end{figure*}

\begin{figure*}
\centering
\includegraphics[width=1\textwidth]{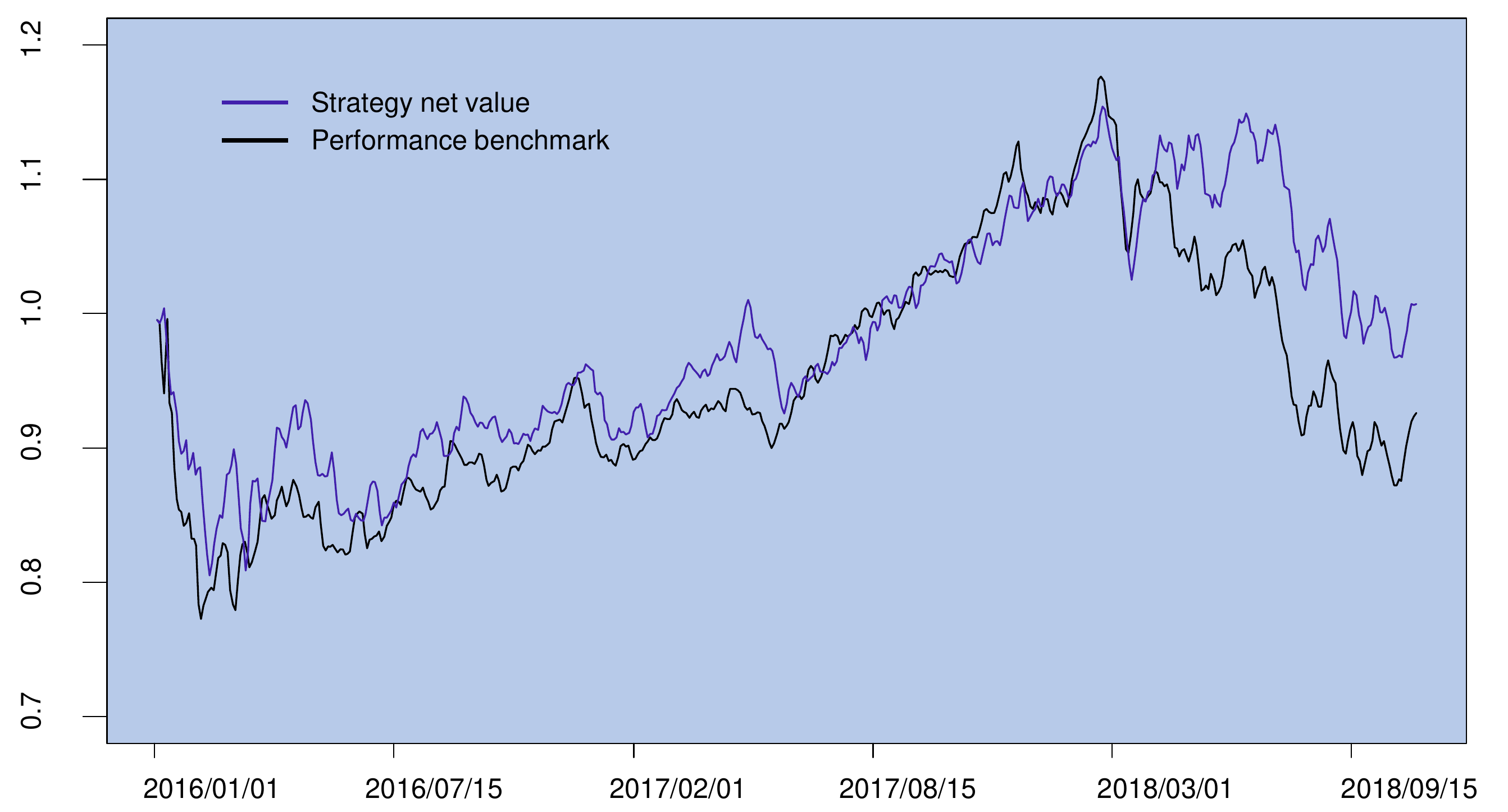}
\caption{Using GARCH model(SVM only)}
\label{imgGARCH and SVM}
\end{figure*}

\begin{figure*}
\centering
\includegraphics[width=1\textwidth]{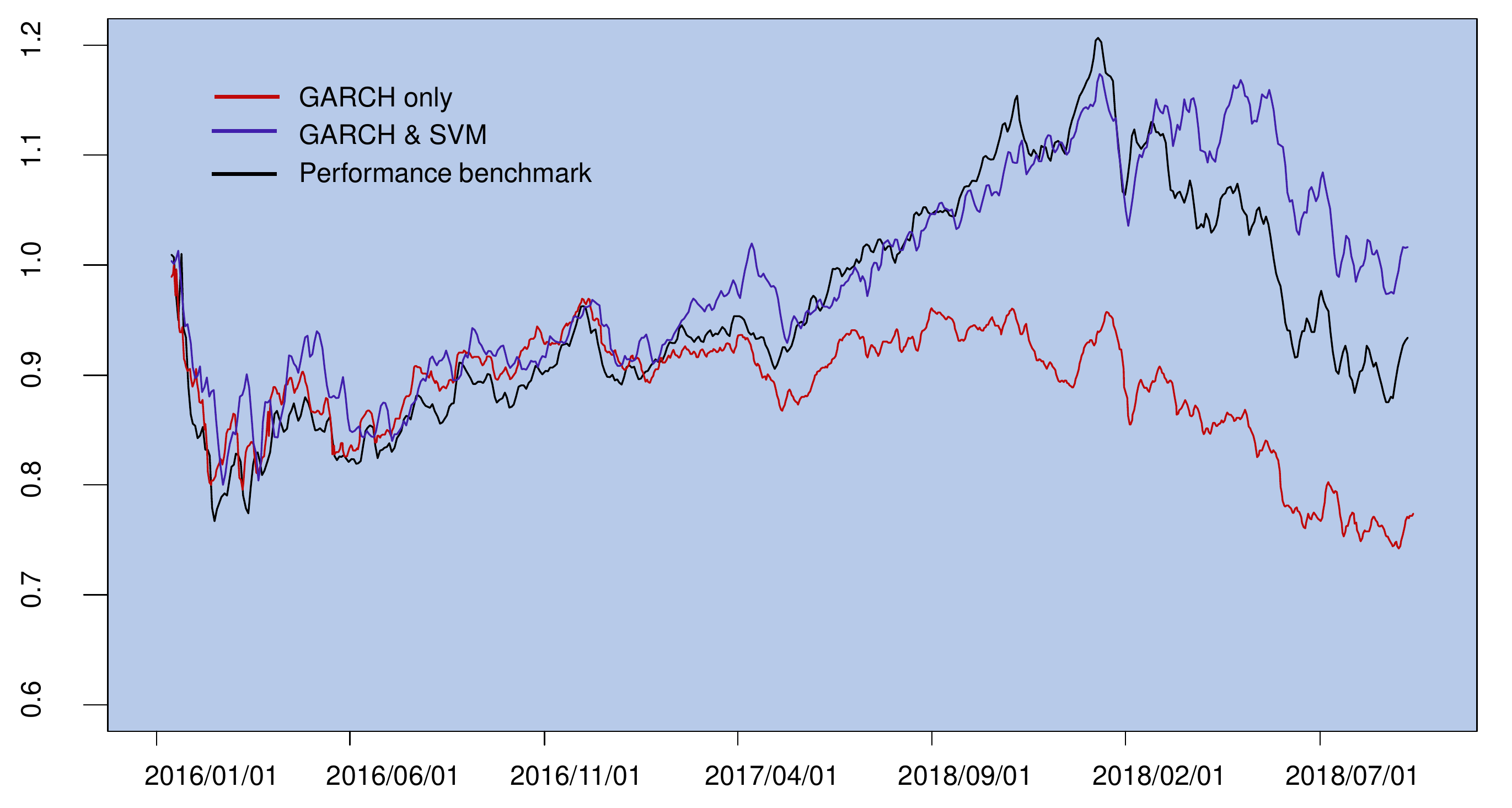}
\caption{Effect comparison(without $VPIN$)}
\label{imgcomparison without $VPIN$}
\end{figure*}

\begin{figure*}
\centering
\includegraphics[width=1\textwidth]{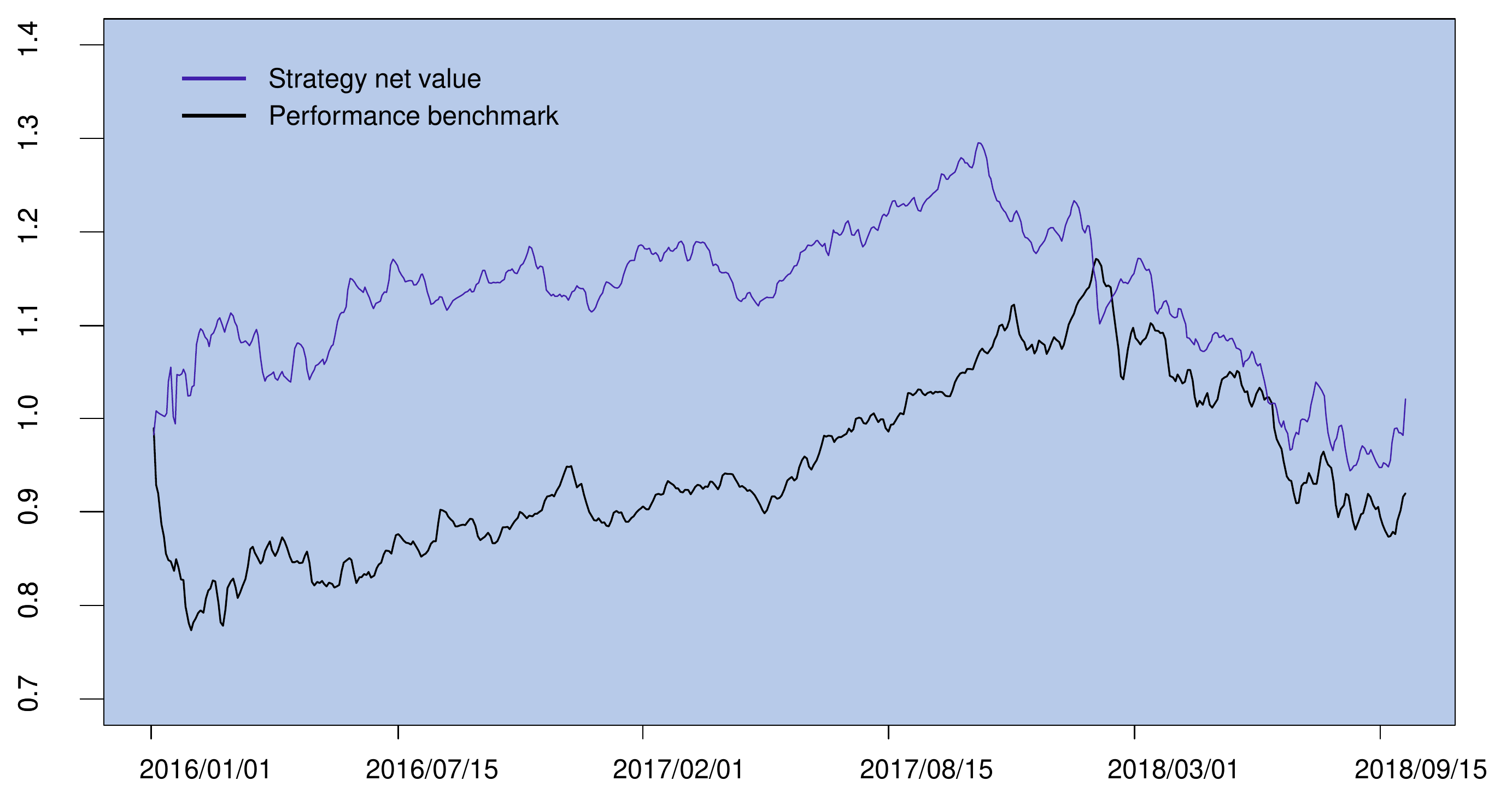}
\caption{Using GARCH model($VPIN$ only)}
\label{imgGARCH and $VPIN$}
\end{figure*}

\begin{figure*}
\centering
\includegraphics[width=1\textwidth]{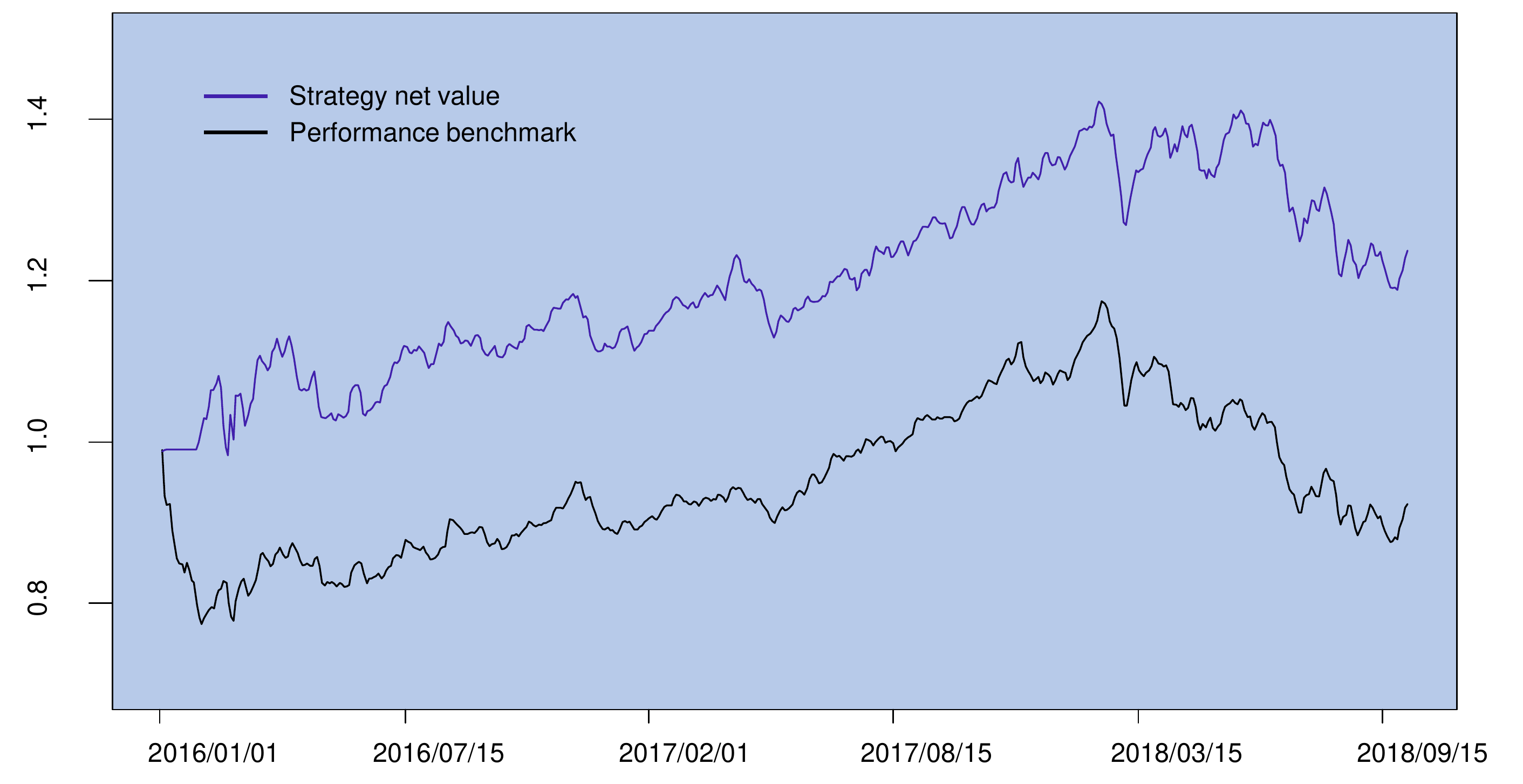}
\caption{Using GARCH model($VPIN$ and SVM)}
\label{imgGARCH, $VPIN$ and SVM}
\end{figure*}

\begin{figure*}
\centering
\includegraphics[width=1\textwidth]{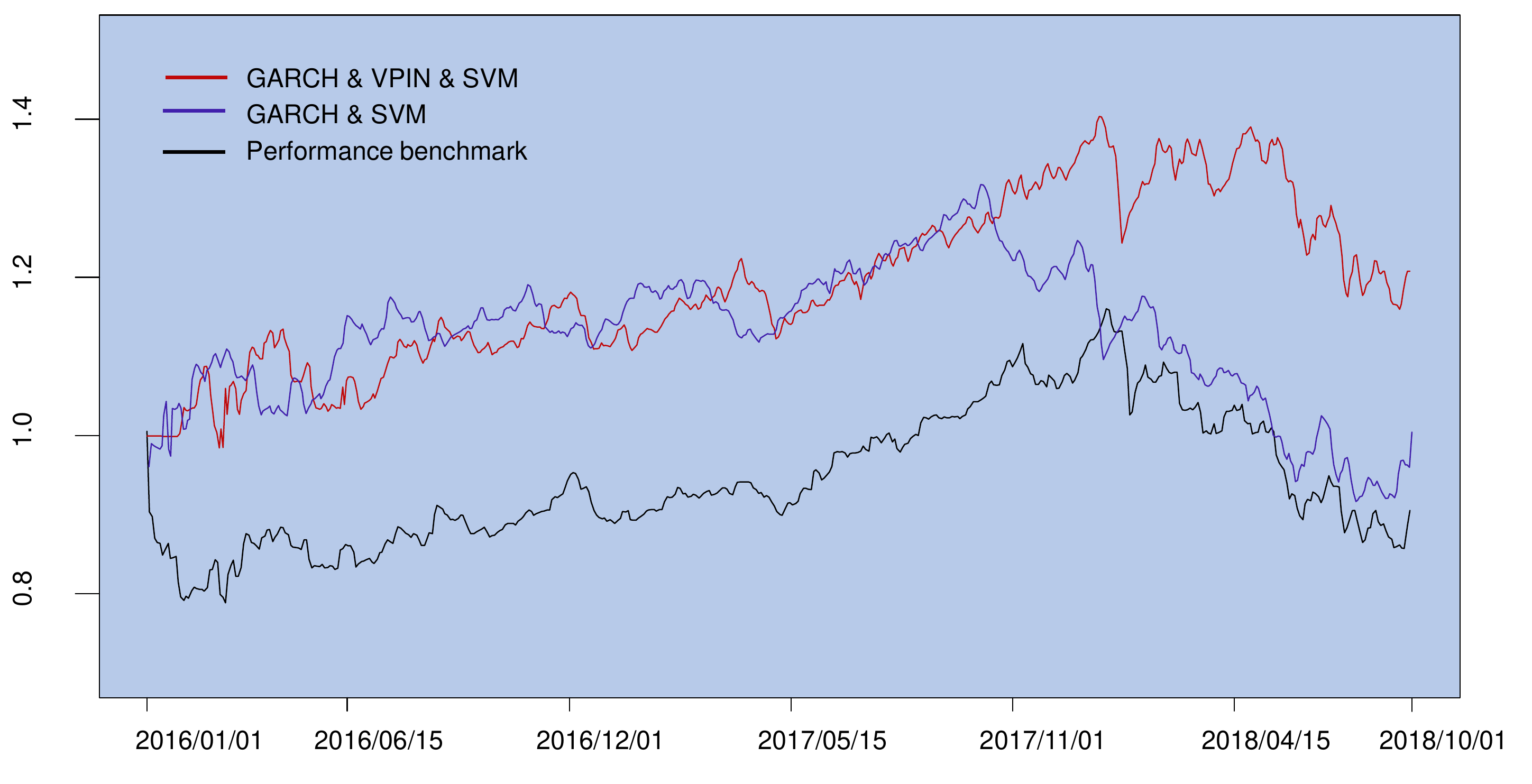}
\caption{Effect comparison(Combined with $VPIN$)}
\label{imgcomparison}
\end{figure*}

\begin{table*}
\centering
\begin{tabular}{lllll}
\hline
& G & G+S & G+V & G+V+S \\ \hline
\begin{tabular}[c]{@{}l@{}}total\\ returns\end{tabular} & -25.85\% & 0.28\% & 3.17\% & 24.23\% \\
\begin{tabular}[c]{@{}l@{}}total\\ annualized returns\end{tabular} & -10.55\% & 0.11\% & 1.17\% & 8.42\% \\
\begin{tabular}[c]{@{}l@{}}relative\\ rate for return\end{tabular} & -18.02\% & 8.11\% & 11.00\% & 32.06\% \\
Alpha & -7.66\% & 3.15\% & 2.49\% & 9.88\% \\
Beta & 0.983 & 1.008 & 0.722 & 0.744 \\
\begin{tabular}[c]{@{}l@{}}max\\ drawdown\end{tabular} & -32.01\% & -21.74\% & -29.98\% & -17.70\% \\
Sharpe & -0.693 & -0.141 & -0.111 & 0.310 \\ \hline
\end{tabular}
\caption{Comparison of strategy evaluation indicators}
\label{tabindicators comparison}
\end{table*}

The figure \ref{imgGARCH only} \ref{imgGARCH and SVM} \ref{imgcomparison without $VPIN$} show the strategy performance without $VPIN$, and the figure \ref{imgGARCH and $VPIN$} \ref{imgGARCH, $VPIN$ and SVM} \ref{imgcomparison} shows the strategy performance combined with $VPIN$. And in \ref{tabindicators comparison}, it shows some indicators which evaluate profitability and stability of the strategy.

Through the previous discussion, this paper improves the customary trading strategy. After considering the liquidity risk and the volatility loss, we have got a strategy that far exceeds the industry benchmark. Even if we do not classify through SVM, with the combining of the GARCH model and $VPIN$ method, the strategy also exceeds the industry benchmark, indicating that the preceding discussion has a significant effect on the improvement of the strategy. Moreover, this paper compares the performance, whether adding the above methods. After the combination of the $VPIN$ method and the GARCH model, although the performance after adding SVM in some time intervals is inferior to the performance of not adding SVM, but consider the $\beta$ and medium- and long-term benefits, combining SVM does have some advantages.

As a result of these, to control risk, we would modify the GARCH threshold with the parameter of SVM to cope with the challenge faced by traditional algorithms. We would further discuss the consequences of high positions. Meanwhile, the $VPIN$ method proved not sensitive to 500 milliseconds of UHF data because of its noise. Easley et al. divide the "basket" into a large number, to avoid the influence of noise on the evaluation ability of $VPIN$, but also affect the ability to process more exceptional data. Therefore, to employ the prediction capability of the GARCH model, this paper dealt with UHF data, used wavelet analysis to reduce noise, and then improve the $VPIN$ method to make this algorithm more suitable for the trading rules of China's stock index futures.

\section{Model Refinement}
First, this paper chooses wavelet analysis to process UHF data. From the calculation method of $VPIN$ that the $VPIN$ method divides the baskets by volume, which also results in a different value of $VPIN$ if changing the number of baskets. So, this paper also improves the $VPIN$ method of volatility hedging. The key to high-frequency data for processing is noise reduction. The simplest way to do this is to abandon some parts of high-frequency data and reduce the frequency to achieve the denoise signal effect naturally. However, it is not reasonable from a statistical point of view. Correct the deviation of noise fluctuations was widely accepted. Many scholars have made efforts in this direction, such as Moving Average Filter used by Andersen \cite{18}, Kalman Filter used by Owens \cite{19}, These methods have been tested based on other trading market data and have obtained good results. They utilized weighted averaging to denoise. This paper use wavelet analysis to denoise, taking the 500 ms data of March 15, 2019, as an example for denoising.

This paper use Haar wavelet as the basis function, level is 6, and the threshold method is fixed from thr.(unscaled wn), soft threshold. Haar function is:

\begin{equation*}
\Psi(t)=\left\{
\begin{aligned}
1 \; \; \; 0 \leqslant t <1\\
0 \; \; \; otherwise .
\end{aligned}
\right.
\end{equation*}

The data fluctuation comparison before and after denoising is shown in the figure:
\begin{figure}[H]
\centering
\includegraphics[width=1\textwidth]{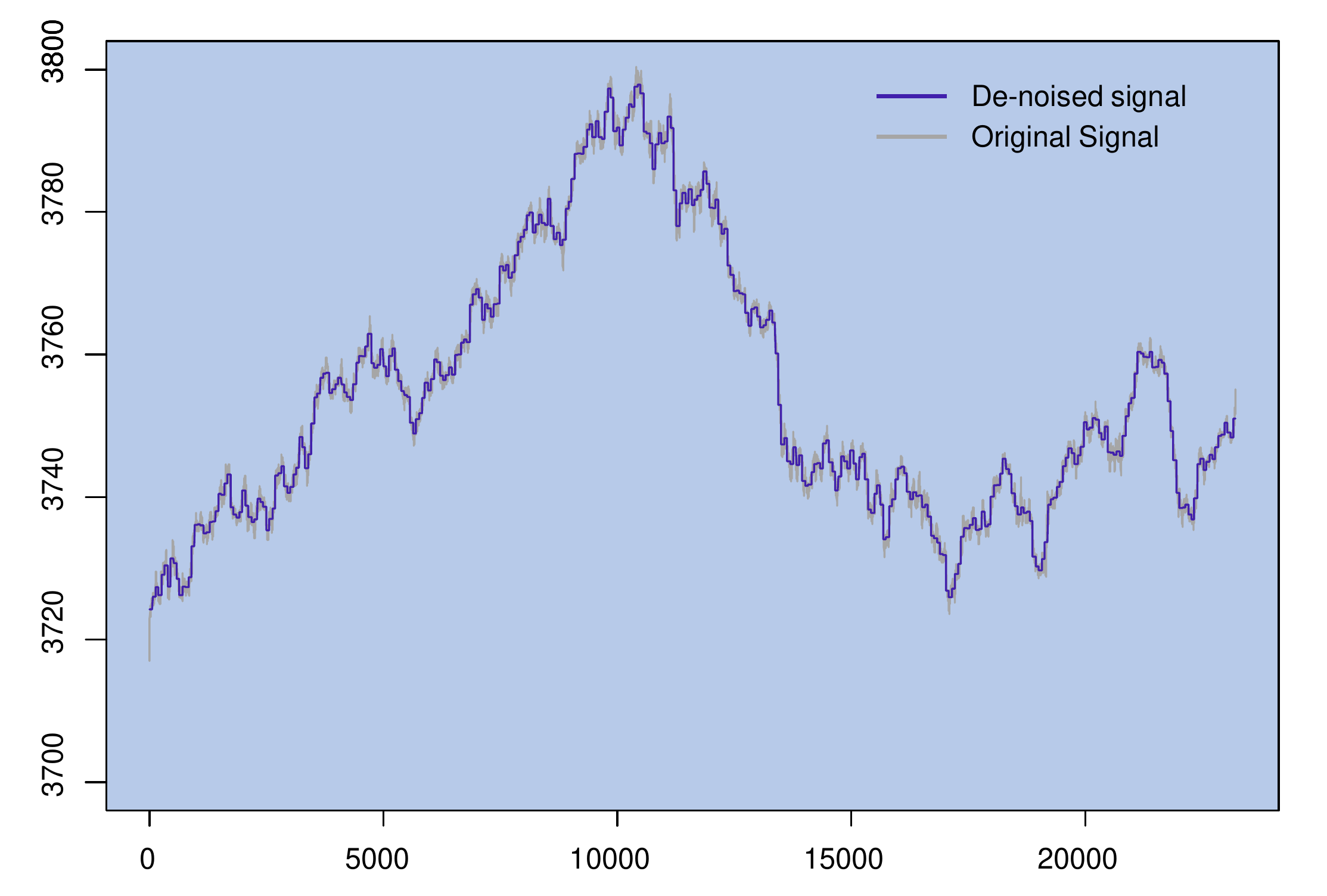}
\caption{comparison before and after denoising}
\end{figure}

This figure illustrated the significance of the denoise algorithm and prepared for further melioration. In the meantime, nevertheless, $VPIN$ does not work well in high-frequency trading, because the $VPIN$ method combines the ask volume and the bid volume in the market, then use the difference between them to predict the volatility changing later. Nevertheless, in high-frequency trading, it is hard to balance the number of baskets and forecasting ability. If the number of $VPIN$ baskets is too hefty, it may lead to a significant decline in the overall forecasting ability. If the number of baskets is too small, it may miss many trading opportunities. So after combining the HAR-RV model, it is a good idea to use different scales volatility to hedge.

The HAR-$VPIN$ model can be written in:
\begin{equation*}
RV_{t,t+H}=\beta_0+\beta_FRV_t^f+\beta_HRV_{t-12,t}^h+\beta_DRV_{t-48,t}^d+\beta_VV_t+\beta_{VPIN}VPIN_t+\varepsilon_{t+H},
\end{equation*}
where $RV_t^f$ is the currently implemented 5 minute volatility, $RV_{t-12,t}^h$ is the currently implemented 1 hour volatility, $RV_{t-48,t}^d$ is the currently implemented one day. Using regression can get the $\beta$ coefficient. This allows for hedging at different scales.

The logic of this model is that different investors have different investment preferences and investment habits. Especially for the Chinese futures market, manual trading is still in the mainstream, so the difference in trading behavior of this group of people will It has a significant influence on price fluctuations. Take into account this feature. It is necessary to analyze multiple time scales.

The HAR-$VPIN$ model has excellent theoretical value. We can perform regression calculation on the HAR-$VPIN$ model to obtain the current short-term and long-term dominant conditions, and adjust the time-window of the previous algorithm according to this situation. Denoised data gets excellent strategic performance. Since the focus of this article is on the idea of the using of high dimensional data, the improvement of the basic model is no longer discussed more.

\section{Conclusion}
In deep learning, the weight values continuously updated through the process shown in the figure,

\begin{figure}[H]
\centering
\includegraphics[width=1\textwidth]{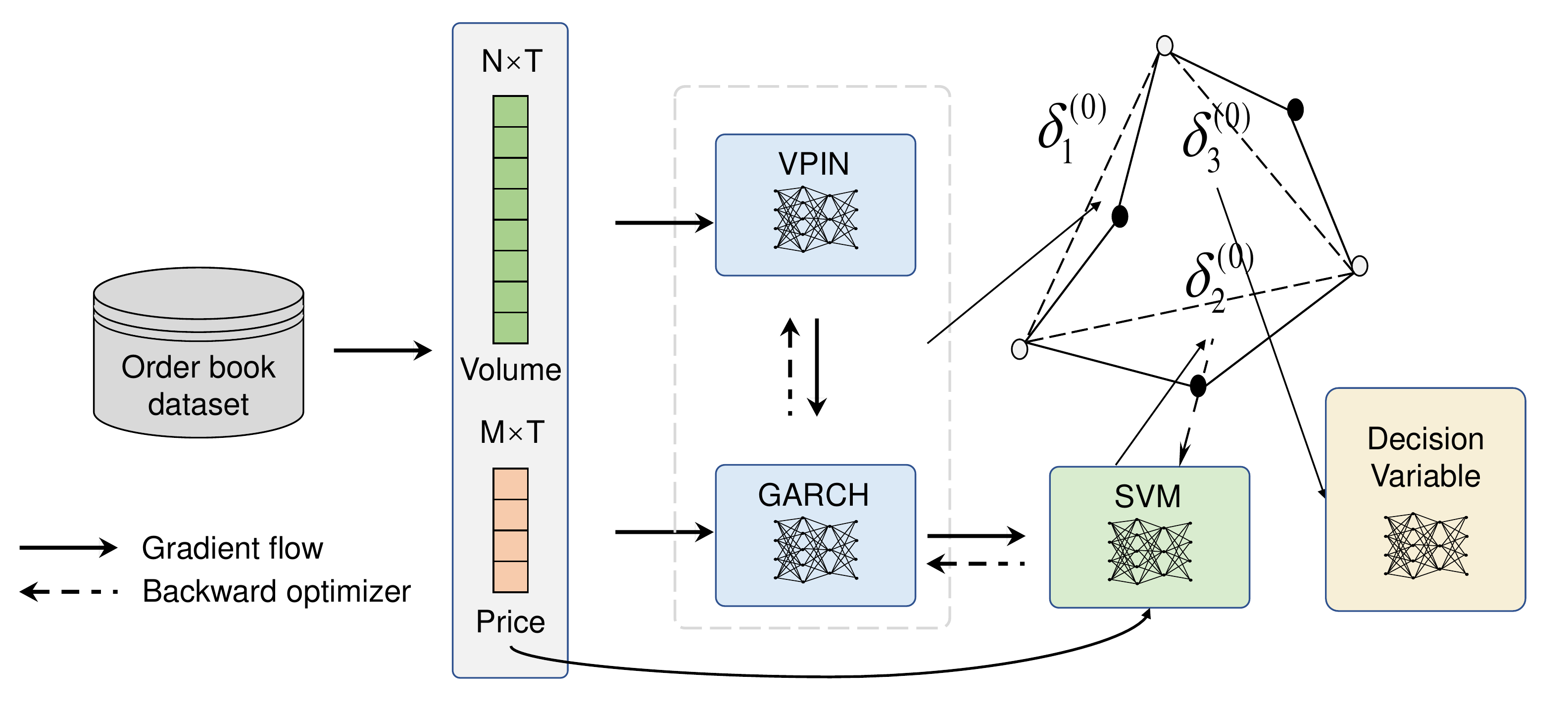}
\caption{Model structure diagram}
\end{figure}
Our model design adjusted through the layers of GARCH, $VPIN$, and SVM. The loss function is obtained based on the maximum yield by adjusting the weight continuously. Stepping on econometrics, creatively employ predictive toolkits in deep learning to complement GARCH, and $VPIN$ provided information beyond the time series. In other words, the activation functions are not necessarily superpositions of simple functions. Instead, when combining the conventional approach with optimization in the stacked layers of networks. The model would improve.

\section*{Funding}

Funded by FDUROP Xi-Yuan Program.

\clearpage

\begin{appendix}
\addappheadtotoc
\appendixpage
\begin{table}[H]
\center
\begin{tabular}{@{}lllll@{}}
\begin{tabular}[c]{@{}l@{}}Null Hypothesis: R has a\\ unit root\end{tabular} & & & & \\
Exogenous: Constant & & & & \\
\begin{tabular}[c]{@{}l@{}}Lag Length: 3 (Automatic\\ - based on SIC, maxlag=55)\end{tabular} & & & & \\ \midrule
& & & t-Statistic & Prob.* \\ \midrule
Augmented Dickey-Fuller test statistic & & & & \\
Test critical values: & 1\% level & & -3.430328 & \\
& 5\% level & & -2.861415 & \\
& 10\% level & & -2.566744 & \\ \bottomrule
\end{tabular}
\caption{ADF test result}
\end{table}

After the first, second and third order lag, the goodness of fit is obviously improved.

\begin{table}[H]
\center
\begin{tabular}{@{}lllll@{}}
\toprule
Variable & Coefficient & Std. Error & t-Statistic & Prob. \\ \midrule
R(-1) & -0.800608 & 0.007903 & -101.3066 & 0.0000 \\
D(R(-1)) & 0.062440 & 0.007046 & 8.861728 & 0.0000 \\
D(R(-2)) & -0.008693 & 0.005910 & -1.471012 & 0.1413 \\
D(R(-3)) & 0.022056 & 0.004754 & 4.639846 & 0.0000 \\
\multicolumn{1}{c}{C} & \multicolumn{1}{c}{-3.00E-06} & \multicolumn{1}{c}{2.91E-06} & \multicolumn{1}{c}{-1.030397} & \multicolumn{1}{c}{0.3028} \\
& & & & \\
\multicolumn{1}{c}{R-squared} & \multicolumn{1}{c}{0.381027} & \multicolumn{1}{c}{Mean dependent var} & \multicolumn{1}{c}{-1.32E-08} & \multicolumn{1}{c}{0.0000} \\
Adjusted R-squared & 0.380971 & S.D. dependent var & 0.000777 & 0.1413 \\
S.E. of regression & 0.000611 & Akaike info criterion & -11.96209 & 0.0000 \\
Sum squared resid & 0.016519 & Schwarz criterion & -11.96111 & 0.3028 \\
\multicolumn{1}{c}{Log likelihood} & \multicolumn{1}{c}{264498.9} & \multicolumn{1}{c}{Hannan-Quinn criter.} & \multicolumn{1}{c}{-11.96179} & \multicolumn{1}{c}{0.0000} \\
\multicolumn{1}{c}{F-statistic} & \multicolumn{1}{c}{6804.754} & \multicolumn{1}{c}{Durbin-Watson stat} & \multicolumn{1}{c}{2.000228} & \multicolumn{1}{c}{0.3028} \\
\multicolumn{1}{c}{Prob(F-statistic)} & \multicolumn{1}{c}{0.000000} & \multicolumn{1}{c}{} & \multicolumn{1}{c}{} & \multicolumn{1}{c}{} \\ \bottomrule
\end{tabular}
\caption{The degree of goodness of fit after the first, second and third order lag}
\end{table}

\indent GARCH(1,1) is $GARCH = C(1) + C(2)*RESID(-1)^2 + C(3)*GARCH(-1)$, It shows that GARCH(1,1) passed the residual test, and the $\alpha+\beta$ value is 0.93, the process of GARCH(1,1) is stable.
\begin{table}[H]
\center
\begin{tabular}{@{}lllll@{}}
\toprule
Variable & Coefficient & Std. Error & z-Statistic & Prob. \\ \midrule
& Variance & Equation & & \\ \midrule
C & 3.35E-08 & 8.37E-10 & 40.08253 & 0.0000 \\
RESID(-1)\textasciicircum{}2 & 0.030665 & 0.001008 & 30.41325 & 0.0000 \\
GARCH(-1) & 0.900759 & 0.002432 & 370.4532 & 0.0000 \\
R-squared & -0.168849 & Mean dependent var & -0.000260 & \\
\multicolumn{1}{c}{Adjusted R-squared} & \multicolumn{1}{c}{-0.168823} & \multicolumn{1}{c}{S.D. dependent var} & \multicolumn{1}{c}{0.000632} & \multicolumn{1}{c}{} \\
S.E. of regression & 0.000683 & Akaike info criterion & -11.79085 & \\
\multicolumn{1}{c}{Sum squared resid} & \multicolumn{1}{c}{0.020646} & \multicolumn{1}{c}{Schwarz criterion} & \multicolumn{1}{c}{-11.79026} & \multicolumn{1}{c}{} \\
Log likelihood & 260734.2 & Hannan-Quinn criter. & -11.79067 & \\
Durbin-Watson stat & 1.293097 & & & \\ \bottomrule
\end{tabular}
\caption{GARCH(1,1) model test}
\end{table}

\indent GARCH(1,2)is:$GARCH = C(1) + C(2)*RESID(-1)^2 + C(3)*GARCH(-1) + C(4)*GARCH(-2)$, it also shows that coefficient of GARCH(1,2) model passes the residual test.

\begin{table}[H]
\center
\begin{tabular}{@{}lllll@{}}
\toprule
Variable & Coefficient & Std. Error & z-Statistic & Prob. \\ \midrule
& Variance & Equation & & \\ \midrule
C & 5.09E-08 & 1.39E-09 & 36.55706 & 0.0000 \\
RESID(-1)\textasciicircum{}2 & 0.047587 & 0.001751 & 27.17560 & 0.0000 \\
GARCH(-1) & 0.359953 & 0.032990 & 10.91088 & 0.0000 \\
GARCH(-2) & 0.488657 & 0.030789 & 15.87141 & 0.0000 \\
\multicolumn{1}{c}{R-squared} & \multicolumn{1}{c}{-0.168849} & \multicolumn{1}{c}{Mean dependent var} & \multicolumn{1}{c}{-0.000260} & \multicolumn{1}{c}{} \\
Adjusted R-squared & -0.168823 & S.D. dependent var & 0.000632 & \\
\multicolumn{1}{c}{S.E. of regression} & \multicolumn{1}{c}{0.000683} & \multicolumn{1}{c}{Akaike info criterion} & \multicolumn{1}{c}{-11.79260} & \multicolumn{1}{c}{} \\
Sum squared resid & 0.020646 & Schwarz criterion & -11.79182 & \\
Log likelihood & 260773.9 & Hannan-Quinn criter. & -11.79236 & \\
Durbin-Watson stat & 1.293097 & & & \\
\bottomrule
\end{tabular}
\caption{GARCH(1,2) model test}
\end{table}

\indent GARCH(2,1) is:$GARCH = C(1) + C(2)*RESID(-1)^2 + C(3)*RESID(-2)^2 + C(4) *GARCH(-1)$, it also shows that coefficient of GARCH(2,1) model passes the residual test.

\begin{table}[H]
\center
\begin{tabular}{@{}lllll@{}}
\toprule
Variable & Coefficient & Std. Error & z-Statistic & Prob. \\ \midrule
& Variance & Equation & & \\ \midrule
C & 1.66E-07 & 3.05E-09 & 54.29680 & 0.0000 \\
RESID(-1)\textasciicircum{}2 & 0.131852 & 0.004239 & 31.10273 & 0.0000 \\
RESID(-2)\textasciicircum{}2 & 0.042368 & 0.004652 & 9.108157 & 0.0000 \\
GARCH(-1) & 0.532821 & 0.008584 & 62.07213 & 0.0000 \\
\multicolumn{1}{c}{R-squared} & \multicolumn{1}{c}{-0.168849} & \multicolumn{1}{c}{Mean dependent var} & \multicolumn{1}{c}{-0.000260} & \multicolumn{1}{c}{} \\
Adjusted R-squared & -0.168823 & S.D. dependent var & 0.000632 & \\
\multicolumn{1}{c}{S.E. of regression} & \multicolumn{1}{c}{0.000683} & \multicolumn{1}{c}{Akaike info criterion} & \multicolumn{1}{c}{-11.77636} & \multicolumn{1}{c}{} \\
Sum squared resid & 0.020646 & Schwarz criterion & -11.77557 & \\
Log likelihood & 260414.7 & Hannan-Quinn criter. & -11.77611 & \\
Durbin-Watson stat & 1.293097 & & & \\
\bottomrule
\end{tabular}
\caption{GARCH(2,1) model test}
\end{table}

\indent TGARCH model is:$GARCH = C(1) + C(2)*RESID(-1)^2 + C(3)*RESID(-1)^2*(RESID(-1)<0) + C(4)*GARCH(-1) $. Through the TGARCH model, we can see that there is a positive leverage effect in the market. So when we set the threshold, we can consider the influence of bad news and good news, and give different impact factors to achieve higher returns.

\begin{table}[H]
\center
\begin{tabular}{@{}lllll@{}}
\toprule
Variable & Coefficient & Std. Error & z-Statistic & Prob. \\ \midrule
& Variance & Equation & & \\ \midrule
C & 3.03E-08 & 8.04E-10 & 37.76715 & 0.0000 \\
RESID(-1)\textasciicircum{}2 & 0.047646 & 0.001880 & 25.34682 & 0.0000 \\
RESID(-1)\textasciicircum{}2*(RESID(-1)\textless{}0) & -0.027411 & 0.001535 & -17.85833 & 0.0000 \\
GARCH(-1) & 0.910709 & 0.002263 & 402.3632 & 0.0000 \\
\multicolumn{1}{c}{R-squared} & \multicolumn{1}{c}{-0.168849} & \multicolumn{1}{c}{Mean dependent var} & \multicolumn{1}{c}{-0.000260} & \multicolumn{1}{c}{} \\
Adjusted R-squared & -0.168823 & S.D. dependent var & 0.000632 & \\
\multicolumn{1}{c}{S.E. of regression} & \multicolumn{1}{c}{0.000683} & \multicolumn{1}{c}{Akaike info criterion} & \multicolumn{1}{c}{-11.79251} & \multicolumn{1}{c}{} \\
Sum squared resid & 0.020646 & Schwarz criterion & -11.79173 & \\
Log likelihood & 260771.8 & Hannan-Quinn criter. & -11.79226 & \\
Durbin-Watson stat & 1.293097 & & &
\\ \bottomrule
\end{tabular}
\caption{TGARCH model test}
\end{table}
\end{appendix}
\end{document}